\documentclass[12pt]{article}%
\usepackage{amssymb}
\usepackage{amsfonts}
\usepackage{amsmath}
\usepackage{graphicx}%
\begin{document}
\begin{titlepage}
\vspace{1.5cm}
\begin{center}

{\bf\large  Charmonium description from a generalized screened potential model }

\date{ }
\vskip 0.70cm
P. Gonz\'{a}lez
\vskip 0.30cm
{ \it Departamento de F\'{\i}sica Te\'orica -IFIC\\
Universitat de València-CSIC \\
E-46100 Burjassot (Valencia), Spain.} \\ ({\small E-mail:
pedro.gonzalez@uv.es})
\end{center}
\vskip 1cm \centerline{\bf Abstract}
A generalized  screened potential model (GSPM), recently developed to study the bottomonium spectrum,
is applied to the calculation of charmonium masses and electromagnetic widths. The presence in the GSPM of more
quark-antiquark bound states than in conventional non screened potential models,
allows for the assignment of GSPM states to cataloged non conventional $J^{++}$ charmonium resonances
as well as for the prediction of new (non cataloged) $J^{++}$ states.
The results obtained seem to indicate that a reasonable overall description of $J^{++}$ charmonium resonances
is feasible.
\end{titlepage}

\section{Introduction\label{SI}}

In a recent paper \cite{Gon14} a new non relativistic quark model for the
description of heavy quark mesons has been developed. The novelty of the
model, called Generalized Screened Potential Model or GSPM, is the
consideration of a lattice motivated quark-antiquark interaction that
implicitly incorporates color screening effects from meson-meson
configurations. When applied to bottomonium a good spectral description of
well established resonances is obtained and a richer high energy spectrum
(bigger number of bound states) than the one resulting from the non-screened
Cornell potential is predicted. However, the current lack of data does not
allow to validate or refute this prediction. In this regard the application of
the model to charmonium could be determinant since a plethora of additional
states, not fitting into the conventional non-screened Cornell potential
framework, has been discovered in the last ten years (see
\cite{PDG14,Ols14,Bra11} and references therein).

In this article we apply the GSPM to charmonium. We extend the observable
analysis beyond the spectral masses to electromagnetic widths, for the model
is suitable for their calculation and there exist data to be compared with. We
do not analyze strong decays since a fully consistent treatment of them within
the GSPM framework (involving the description of mesons containing light
quarks) is a formidable task outside the scope of the present study.

We show that a reasonable description of well established and candidates to
$J^{++}$ resonances is feasible. Moreover the model allows for some definite
predictions about new resonances what might be used in future experimental
searches to further check its validity. Regarding $1^{--}$ states the presence
of overlapping thresholds limits the applicability of the GSPM to spectral
energies quite below the first meson-meson threshold becoming then completely
equivalent to the Cornell model.

The article contents are organized as follows. In Section \ref{SII} a brief
review of the GSPM is presented. In Section \ref{SIII} the model is applied to
the calculation of the charmonium spectrum and electromagnetic widths and the
results are compared to data. A calculation from a non screened Cornell
potential is also shown for comparison. Finally, in Section \ref{SIV} our main
results and conclusions are summarized.

\section{Generalized Screened Potential Model (GSPM)\label{SII}}

The Generalized Screened Potential Model (GSPM) is based on the assumption
that a heavy quark meson description can be attained from the consideration of
effective valence quark degrees of freedom interacting through a potential
that incorporates screening effects from meson-meson configurations.

More precisely, the Generalized Screened Potential that we shall call
$V\left(  r\right)  $ henceforth tries to implement within a quark model
framework the lattice results for the energy of two static color sources
(heavy quark and heavy antiquark) in terms of their distance, $E_{lattice}%
\left(  r\right)  $, when the mixing of the quenched quark-antiquark
configuration with open flavor meson-meson ones is taken into account. In
reference \cite{Bal05} the lattice calculation for the case of one open flavor
meson-meson configuration was performed, the resulting $E_{lattice}\left(
r\right)  $ having a different form below and above the meson-meson threshold.
For the two threshold case an educated guess for $E_{lattice}\left(  r\right)
$ was done (see Fig. 22 in \cite{Bal05}). A simplified generalization of these
lattice results to the many threshold case was proposed in reference
\cite{Gon14} . From it a static quark-antiquark potential, $V\left(  r\right)
,$ was derived by means of a Born-Oppenheimer approximation, say by
subtracting the quark and antiquark masses, $m_{Q}$ and $m_{\overline{Q}}$,
from the static energy.

\bigskip

Explicitly, by calling $M_{T_{i}}$ with $i\geq1$ the masses of the physical
meson-meson thresholds, $T_{i}$, with a given set of quantum numbers
$I(J^{PC})$, and defining $M_{T_{0}}\equiv0$ for a unified notation (note that
$T_{0}$ does not correspond to any physical meson-meson threshold), the form
of $V\left(  r\right)  $ in the different energy regions (specified as energy
interval subindices) reads:%

\begin{equation}
V_{\left[  M_{T_{0}},M_{T_{1}}\right]  }(r)=\left\{
\begin{array}
[c]{c}%
\sigma r-\frac{\chi}{r}\text{ \ \ \ \ \ \ \ \ \ \ \ \ \ \ \ \ \ \ \ \ \ \ }%
r\leq r_{T_{1}}\\
\\
M_{T_{1}}-m_{Q}-m_{\overline{Q}}\text{ \ \ \ \ \ \ \ \ \ }r\geq r_{T_{1}}%
\end{array}
\right.  \label{pot1}%
\end{equation}
and%
\begin{equation}
V_{\left[  M_{T_{j-1}},M_{T_{j}}\right]  }(r)=\left\{
\begin{array}
[c]{c}%
M_{T_{j-1}}-m_{Q}-m_{\overline{Q}}\text{ \ \ \ \ \ }r\leq r_{T_{j-1}}\\
\\
\sigma r-\frac{\chi}{r}\text{\ \ \ \ \ \ \ \ \ \ \ }r_{T_{j-1}}\leq r\leq
r_{T_{j}}\\
\\
M_{T_{j}}-m_{Q}-m_{\overline{Q}}\text{ \ \ \ \ \ \ \ \ \ }r\geq r_{T_{j}}%
\end{array}
\right.  \label{pot12}%
\end{equation}
for $j>1,$ with the crossing radii $r_{T_{i}}$ $\left(  i\geq1\right)  $
defined by
\begin{equation}
\sigma r_{T_{i}}-\frac{\chi}{r_{T_{i}}}=M_{T_{i}}-m_{Q}-m_{\overline{Q}}
\label{rti}%
\end{equation}
with the parameters $\sigma$ and $\chi$ standing for the string tension and
the color coulomb strength respectively.

Thus $V\left(  r\right)  $ has in each energy region between neighbor
thresholds a Cornell form but modulated at short and long distances by these thresholds.

\bigskip

\begin{figure}
[ptb]
\begin{center}
\includegraphics[
height=2.3808in,
width=3.659in
]%
{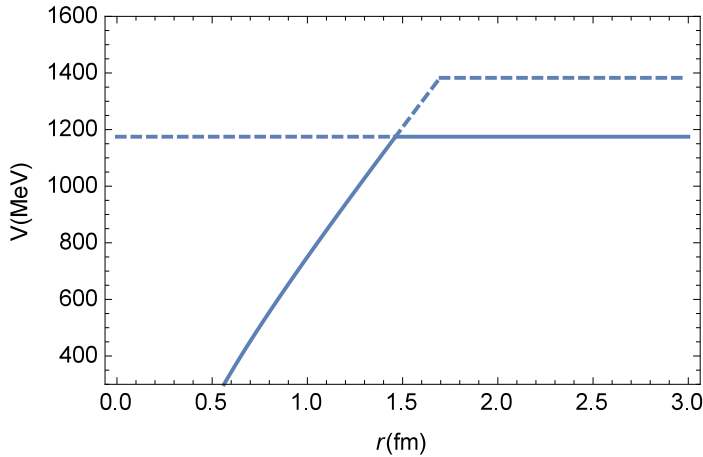}%
\caption{Generalized screened potential $V(r).$ The solid (dashed) line
indicates the potential in the first (second) energy region for $0^{+}%
(1^{++})$ $c\overline{c}$ states with $m_{c}=1348.6$ MeV, $\sigma=850$ MeV/fm,
$\chi=100$ MeV.fm, $M_{T_{1}}=3872$ MeV and $M_{T_{2}}=4080$ MeV (values of
the parameters and threshold masses from Section \ref{SIII}).}%
\label{Figcharm850pot}%
\end{center}
\end{figure}

Thus for example in Fig. \ref{Figcharm850pot} the form of $V\left(  r\right)
$ in the\ first and second energy regions is drawn for $c\overline{c}$ states
with $I^{G}(J^{PC})=0^{+}(1^{++})$ quantum numbers, whose first threshold
$T_{1}$ corresponds to $D\overline{D^{\ast}}$ and its second threshold $T_{2}$
to $D_{s}^{+}D_{s}^{\ast-}$.

%

\bigskip

Let us remark that $V\left(  r\right)  $ is an energy dependent potential in
the sense that its form differs in the different energy regions delimited by
the thresholds. Actually this is the essential difference with other screened
potential models which have been also employed for the description of
charmonium \cite{Li09,Gon09}.

It is also important to emphasize that in any energy region the potential is
strictly confining in the sense that only bound states are obtained as
solutions of the Schr\"{o}dinger equation in such energy region (see next Section).

\section{Charmonium \label{SIII}}

From the defined $V\left(  r\right)  $, charmonium states in the energy region
$\left[  M_{T_{i-1}},M_{T_{i}}\right]  $, characterized by a definite set of
quantum numbers $I^{G}(J^{PC}),$ are obtained by solving the Schr\"{o}dinger
equation for $V_{\left[  M_{T_{i-1}},M_{T_{i}}\right]  }(r)$.

In order to get the solutions we previously fix the values of the parameters
of the model and list the open charm meson-meson threshold masses to be
considered. Then we detail the calculation of the spectrum for a particular
case before giving the general results. Next we assign calculated states to
experimental resonances and use the corresponding wave functions to evaluate
electromagnetic widths.

\subsection{Parameters\label{SIIIA}}

Let us realize that for energies quite below the first corresponding
thresholds the potential $V\left(  r\right)  $ is almost completely equivalent
to a Cornell one:
\begin{equation}
V_{Cor}(r)\equiv\sigma r-\frac{\chi}{r}\text{ \ \ \ }\left(  r:0\rightarrow
\infty\right)  \label{Cor}%
\end{equation}
But the use of the conventional Cornell potential teaches us that the
charm-anticharm $\left(  c\overline{c}\right)  $ system may be a relativistic
one \cite{Eic80}. This makes debatable the application of the GSPM to
charmonium. In the spirit of quark model calculations we shall assume that the
effectiveness of the parameters (quark mass, string tension and coulomb
strength) may be appropriately taking into account, at least in part,
relativistic corrections. We can also invoke the effectiveness of the
parameters regarding additional contributions from light quark-antiquark pairs
apart from the implicitly considered meson-meson configurations.

\bigskip

Let us also note that $i)$ no threshold widths have been considered and $ii)$
the accumulative interacting effect from different meson-meson configurations
with the same threshold mass has not been implemented. (Indeed if a
meson-meson configuration gives rise to a screening of the quark and antiquark
color charges for a given energy then a reinforcement of the screening is
expected when more meson-meson configurations with the same threshold mass are available.)

Hence we shall restrict the application of the model to those energy regions
involving non degenerate, isolated\ (in the sense of not having any
significant experimental overlap due to their widths) thresholds.

Even so the model may be too simplistic for an accurate description of real
mesons. On the one hand $V(r)$ does not contain spin dependent terms that we
know may give significant contributions to the masses of the lower spectral
states (see for example \cite{GI85}). On the other hand the effect of any
threshold has been approximated by an abrupt (instead of a physically soft)
change in the potential at the crossing radii. Moreover $SU\left(  3\right)  $
flavor symmetry has been considered when the same effect (flattening of the
potential from the crossing radii) from thresholds with $s\overline{s}$,
$u\overline{u}$ or $d\overline{d}$ content has been implemented despite the
fact that the probability of formation for each of these pairs may be different.

Keeping in mind these possible shortcomings we shall try to show that such a
simple model could provide us with some insight onto the dominant dynamic
mechanisms governing the charmonium structure.

As we are dealing with a spin independent potential we shall compare as usual
the calculated $s-$ wave state masses with spin-triplet data, the $p-$ wave
state masses with the centroids obtained from data and the $d-$ wave states
with the few existing experimental candidates.

Aiming at a joint description of charmonium and bottomonium we shall use for
both the same values for the parameters of the potential. From \cite{Gon14} we
have $\sigma=850$ MeV/fm and\ $\chi=100$ MeV.fm. Let us realize that this
string tension value $\sqrt{\sigma}=410$ MeV is within the interval usually
accepted for it from phenomenology (see for instance \cite{Bal01}). As for the
Coulomb strength $\chi$ its value corresponds to a strong quark-gluon-quark
coupling $\alpha_{s}=\frac{3\chi}{4\hbar}\simeq0.38,$ in agreement with the
value derived from QCD from the fine structure splitting of $1p$ states in
charmonium \cite{Bad99}. Regarding the remaining parameter of the model
$m_{c}$ we fix its value to get a reasonable overall fit to the spectrum.

Thus the set of parameters that will be used henceforth is%
\begin{equation}%
\begin{array}
[c]{c}%
\sigma=850\text{ MeV/fm}\\
\chi=100\text{ MeV.fm}\\
m_{c}=1348.6\text{ MeV}%
\end{array}
\label{valpar}%
\end{equation}
where the value of the charm mass $m_{c}=1348.6$ MeV has been fine tuned to
fit the mass of the well established non conventional charmonium state
$X(3872)$ since this resonance may be naturally described in the GSPM as
explained later on.

It is important to remark that due to the effective character of the
parameters a better overall spectral fit (differences from calculated masses
to data of $35$ MeV at most) could be achieved by choosing for example
$\sigma=750$ MeV/fm, $\chi=100$ MeV.fm and $m_{c}=1407.8$ MeV. However as this
new fit makes no difference at all in the resulting number of spectral states
we prefer to maintain the same potential description as in bottomonium.

\subsection{$I(J^{PC})$ Thresholds\label{SIIIB}}

In order to apply the GSPM to a particular set of charmonium $\left(
I=0\right)  $ states with definite $J^{PC}$ we need the masses $M_{T_{i}}$ for
open charm meson - meson thresholds coupling to these quantum numbers. From
these masses the crossing radii $r_{T_{i}}$ are immediately calculated from
(\ref{rti}).

Let us realize that the static approach we follow to build the potential
implies that the heavy quark and antiquark or, quite equivalently, the two
charmed mesons forming the threshold, are in a relative $S-$ wave so that the
threshold mass is just the sum of the masses of the mesons.

The list of known thresholds, their masses and the corresponding crossing
radii appear in Tables~\ref{tab1++} and \ref{tab1--} where a simplified
notation has been used: a threshold has been denoted by the first meson-meson
component entering in the $I\left(  J^{PC}\right)  $ linear combination. Thus,
the first $0(1^{++})$ threshold in Table~\ref{tab1++}, $D^{0}\overline
{D^{\ast}}^{0}(2007)$ denotes $(D^{0}\overline{D^{\ast}}^{0}(2007),D^{+}%
D^{\ast}(2010)^{-})_{I=0,J^{P}=1^{+}}+c.c.$ where $c.c.$ stands for the charge conjugate.

\begin{center}
\begin{table}[ptb]%
\begin{tabular}
[c]{cccccc}%
$I(J^{PC})$ & $T_{i}$ & Charmonium Thresholds & $\left(  J_{1}^{P},J_{2}%
^{P}\right)  $ & $%
\begin{array}
[c]{c}%
M_{T_{i}}\\
\text{(MeV)}%
\end{array}
$ & $%
\begin{array}
[c]{c}%
r_{T_{i}}\\
\text{(fm)}%
\end{array}
$\\\hline
&  &  &  &  & \\
$0(0^{++})$ &  &  &  &  & \\
& $T_{1}$ & $D^{0}\overline{D}^{0}$ & $\left(  0^{-},0^{-}\right)  $ & $3730$
& $1.31$\\
& $T_{2}$ & $D_{s}^{+}D_{s}^{-}$ & $\left(  0^{-},0^{-}\right)  $ & $3937$ &
$1.54$\\
& $T_{3}$ & $D^{\ast0}(2007)\overline{D^{\ast}}^{0}(2007)$ & $\left(
1^{-},1^{-}\right)  $ & $4014$ & $1.62$\\
& $T_{4}$ & $D_{s}^{\ast+}D_{s}^{\ast-}$ & $\left(  1^{-},1^{-}\right)  $ &
$4224$ & $1.86$\\
& $T_{5}$ & $D^{0}\overline{D}^{0}\left(  2550\right)  $ & $\left(
0^{-},0^{-}\right)  $ & $4405$ & $2.07$\\
&  &  &  &  & \\
&  &  &  &  & \\
$0(1^{++})$ &  &  &  &  & \\
& $T_{1}$ & $D^{0}\overline{D^{\ast}}^{0}(2007)$ & $\left(  0^{-}%
,1^{-}\right)  $ & $3872$ & $1.46$\\
& $T_{2}$ & $D_{s}^{+}D_{s}^{\ast-}$ & $\left(  0^{-},1^{-}\right)  $ & $4080$
& $1.70$\\
&  &  &  &  & \\
$0(2^{++})$ &  &  &  &  & \\
& $T_{1}$ & $D^{\ast}(2007)^{0}\overline{D^{\ast}}^{0}(2007)$ & $\left(
1^{-},1^{-}\right)  $ & $4014$ & $1.62$\\
& $T_{2}$ & $D_{s}^{\ast+}D_{s}^{\ast-}$ & $\left(  1^{-},1^{-}\right)  $ &
$4224$ & $1.86$\\
&  &  &  &  &
\end{tabular}
\caption{Open charm meson-meson thresholds for $0\left(  J^{++}\right)  $
charmonium states. $J_{1}^{P}$ and $J_{2}^{P}$ stand for the angular momenta
of the mesons forming the threshold. Threshold masses $\left(  M_{T_{i}%
}\right)  $ obtained from the charmed and charmed strange meson masses quoted
in \cite{PDG14}. Crossing radii $\left(  r_{T_{i}}\right)  $ calculated from
(\ref{rti}).}%
\label{tab1++}%
\end{table}\begin{table}[ptbptb]%
\begin{tabular}
[c]{cccccc}%
$I\left(  J^{PC}\right)  $ & $T_{i}$ & Charmonium Thresholds & $\left(
J_{1}^{P},J_{2}^{P}\right)  $ & $%
\begin{array}
[c]{c}%
M_{T_{i}}\\
\text{(MeV)}%
\end{array}
$ & $%
\begin{array}
[c]{c}%
r_{T_{i}}\\
\text{(fm)}%
\end{array}
$\\
&  &  &  &  & \\
$0\left(  1^{--}\right)  $ &  &  &  &  & \\
& $T_{1}$ & $%
\begin{array}
[c]{c}%
D^{0}\overline{D_{1}}^{0}(2420)\\
D^{0}\overline{D_{1}}^{0}(2430)
\end{array}
$ & $\left(  0^{-},1^{+}\right)  $ & $4287$ & $1.93$\\
&  &  &  &  & \\
& $T_{2}$ & $D^{\ast}(2007)^{0}\overline{D_{0}^{\ast}}^{0}(2400)$ & $\left(
1^{-},0^{+}\right)  $ & $4325$ & $1.98$\\
&  &  &  &  & \\
& $T_{3}$ & $%
\begin{array}
[c]{c}%
D^{\ast}(2007)^{0}\overline{D_{1}}^{0}(2420)\\
D^{\ast}(2007)^{0}\overline{D_{1}}^{0}(2430)\\
D_{s}^{+}D_{s1}(2460)^{-}\\
D_{s}^{\ast+}D_{s0}^{\ast}(2317)^{-}%
\end{array}
$ & $%
\begin{array}
[c]{c}%
\left(  1^{-},1^{+}\right) \\
\left(  1^{-},1^{+}\right) \\
\left(  0^{-},1^{+}\right) \\
\left(  1^{-},0^{+}\right)
\end{array}
$ & $4429$ & $2.09$\\
&  &  &  &  & \\
& $T_{4}$ & $D^{\ast}(2007)^{0}\overline{D_{2}^{\ast}}^{0}(2460)$ & $\left(
1^{-},2^{+}\right)  $ & $4470$ & $2.14$\\
& $T_{5}$ & $D_{s}^{+}D_{s1}(2536)^{-}$ & $\left(  0^{-},1^{+}\right)  $ &
$4504$ & $2.18$\\
& $T_{6}$ & $D_{s}^{\ast+}D_{s1}(2460)^{-}$ & $\left(  1^{-},1^{+}\right)  $ &
$4572$ & $2.26$\\
& $T_{7}$ & $D_{s}^{\ast+}D_{s1}(2536)^{-}$ & $\left(  1^{-},1^{+}\right)  $ &
$4648$ & $2.35$\\
& $T_{8}$ & $D_{s}^{\ast+}D_{s2}^{\ast}(2573)^{-}$ & $\left(  1^{-}%
,2^{+}\right)  $ & $4685$ & $2.39$\\
&  &  &  &  &
\end{tabular}
\caption{Open charm meson-meson thresholds for $0\left(  1^{--}\right)  $
charmonium states. $J_{1}^{P}$ and $J_{2}^{P}$ stand for the angular momenta
of the mesons forming the threshold. Threshold masses $\left(  M_{T_{i}%
}\right)  $ obtained from the charmed and charmed strange meson masses quoted
in \cite{PDG14}. Crossing radii $\left(  r_{T_{i}}\right)  $ calculated from
(\ref{rti}).}%
\label{tab1--}%
\end{table}
\end{center}

We have used isospin symmetry to construct thresholds with well defined
isospin. This means that we are neglecting the mass differences between the
electrically neutral and charged members of the same isospin multiplet, for
example $D^{0}$ and $D^{\pm}$ with PDG quoted masses \cite{PDG14}
$1864.91\pm0.17$ and $1869.5\pm0.4$ respectively or $D^{\ast}(2007)^{0}$ and
$D^{\ast}(2010)^{-}$ with quoted masses $2006.98\pm0.15$ and $2010.21\pm0.13$
respectively. For the calculation of the threshold masses we have used the
lower mass value in any isospin multiplet ($1865$ MeV and $2007$ MeV in the
examples just mentioned).

\bigskip

Regarding the $C$ parity, for a threshold formed by two mesons $\mathfrak{M}%
_{1}$ and $\mathfrak{M}_{2}$ we can construct the combinations $\left(
\mathfrak{M}_{1}\mathfrak{M}_{2}\pm c.c.\right)  $ with $C$ parity $+$ and $-$
respectively. Notice though that if $\mathfrak{M}_{2}=\overline{\mathfrak{M}%
_{1}}$ then, as the two mesons are in $S-$ wave, we have $\overline
{\mathfrak{M}_{1}}\mathfrak{M}_{1}=\left(  -\right)  ^{j_{1}+j_{1}%
-j}\mathfrak{M}_{1}\overline{\mathfrak{M}_{1}}$ where $j_{1}$ stands for the
spin of $\mathfrak{M}_{1}$ and $j$ for the total spin of the threshold.
Therefore only one combination in $\mathfrak{M}_{1}\overline{\mathfrak{M}_{1}%
}\pm$ $c.c.$ is allowed for a given value of $j$ (the other vanishes)$.$ For
example the $I=0$ threshold $D^{\ast}\overline{D^{\ast}}$ with $j_{1}=1$ has
only positive $C$ parity when coupled to $j=0,2$ (and only negative $C$ parity
when coupled to $j=1$).

\bigskip

A look at Table~\ref{tab1++} makes clear that for $0\left(  J^{++}\right)  $
states only non degenerate isolated thresholds are present. Therefore the
GSPM\ can be safely applied. On the contrary for $0\left(  1^{--}\right)  $
states, Table~\ref{tab1--}, there are degenerate ($T_{1}$ and $T_{3}$) and
overlapping thresholds (for instance $T_{1}$ and $T_{3}$ overlap with $T_{2}$
due to the large width ($267$ MeV) of $\overline{D_{0}^{\ast}}^{0}(2400)$).
Therefore, at its present stage the GSPM can only be consistently applied to
$0\left(  1^{--}\right)  $ states quite below the first threshold. In
consequence we shall restrict our study in this case to the first energy region.

\subsection{Spectrum\label{SIIIC}}

Charmonium $(c\overline{c})$ states are obtained by solving the
Schr\"{o}dinger equation for $V(r).$ As in any energy region it is a radial
potential we use spectroscopic notation $k\equiv nl,$ in terms of the radial,
$n,$ and orbital angular momentum, $l,$ quantum numbers, to denote its bound
states. Thus in the energy region $\left[  M_{T_{i-1}},M_{T_{i}}\right]  $ we
have%
\begin{align}
&  \left(  \mathcal{T}+V_{\left[  M_{T_{i-1}},M_{T_{i}}\right]  }\right)
\left\vert (c\overline{c})_{k_{\left[  T_{i-1},T_{i}\right]  }}\right\rangle
\label{SchEQM}\\
&  =M_{k_{\left[  T_{i-1},T_{i}\right]  }}\left\vert (c\overline
{c})_{k_{\left[  T_{i-1},T_{i}\right]  }}\right\rangle \nonumber
\end{align}
where $\mathcal{T}$\ stands for the kinetic energy operator, $\left\vert
(c\overline{c})_{k_{\left[  T_{i-1},T_{i}\right]  }}\right\rangle $ for the
bound states and $M_{k_{\left[  T_{i-1},T_{i}\right]  }}$ for their masses.

\bigskip

Let us consider for example the $0^{+}(1^{++})$ spectral states. In the first
energy region the potential $V(r)$ has the form $V_{\left[  M_{T_{0}}%
,M_{T_{1}}\right]  }(r),$ given by (\ref{pot1}) (solid line in Fig.
\ref{Figcharm850pot})
\begin{equation}
V_{\left[  0,3872\right]  }(r)=\left\{
\begin{array}
[c]{c}%
\sigma r-\frac{\chi}{r}\text{ \ \ \ \ \ \ \ \ \ \ \ \ \ \ \ \ }r\leq1.46\text{
fm}\\
\\
1174.8\text{ MeV\ \ \ \ \ \ \ \ \ \ \ \ \ \ }r\geq1.46\text{ fm}%
\end{array}
\right.  \label{VT0T1}%
\end{equation}
where $M_{T_{1}}$ and $r_{T_{1}}$ have been taken from Table \ref{tab1++} and
the values of the parameters $\left(  \sigma,\chi,m_{c}\right)  $ are given by
(\ref{valpar}).

\bigskip

By solving the Schr\"{o}dinger equation for $V_{\left[  0,3872\right]  }(r)$
we get the GSPM $0^{+}(1^{++})$ spectrum in the first energy region $\left[
M_{T_{0}}=0,M_{T_{1}}=3872\text{ MeV}\right]  $. It has two bound states
states, $1p_{\left[  T_{0},T_{1}\right]  }$ and $2p_{\left[  T_{0}%
,T_{1}\right]  },$ whose masses $M_{k_{\left[  T_{0},T_{1}\right]  }}$
generically denoted by $M_{GSPM}$ are listed in Table~\ref{tabsp1++}.

\bigskip

In the second energy region, $\left[  M_{T_{1}}=3872\text{ MeV},M_{T_{2}%
}=4080\text{ MeV }\right]  ,$ the potential $V(r)$ has the form $V_{\left[
M_{T_{1}},M_{T_{2}}\right]  }(r),$ given by (\ref{pot12}) (dashed line in Fig.
\ref{Figcharm850pot}):
\begin{equation}
V_{\left[  3872,4080\right]  }(r)=\left\{
\begin{array}
[c]{c}%
1174.8\text{ MeV\ \ \ \ \ \ \ \ }r\leq1.46\text{ fm}\\
\\
\sigma r-\frac{\chi}{r}\text{\ \ \ \ \ }1.46\text{ fm}\leq r\leq1.70\text{
fm}\\
\\
1382.8\text{ MeV\ \ \ \ \ \ \ \ }r\geq1.70\text{ fm}%
\end{array}
\right.  \label{VT1T2}%
\end{equation}
where the threshold masses and crossing radii are taken from Table
\ref{tab1++}. The spectrum has only one bound state $1p_{\left[  T_{1}%
,T_{2}\right]  }$ whose mass $M_{1p_{\left[  T_{1},T_{2}\right]  }}$
generically denoted by $M_{GSPM}$ is listed in Table~\ref{tabsp1++}.

\begin{center}
\begin{table}[ptb]%
\begin{tabular}
[c]{ccccc}%
$%
\begin{array}
[c]{c}%
c\overline{c}\\
0(1^{++})
\end{array}
$ & $\left[  T_{i-1},T_{i}\right]  $ & $%
\begin{array}
[c]{c}%
\left[  M_{T_{i-1}},M_{T_{i}}\right] \\
\text{MeV}%
\end{array}
$ & $%
\begin{array}
[c]{c}%
\text{GSPM}\\
\text{States}\\
k_{\left[  T_{i-1},T_{i}\right]  }%
\end{array}
$ & $%
\begin{array}
[c]{c}%
M_{GSPM}\\
\text{MeV}%
\end{array}
$\\\hline
&  &  &  & \\
& $\left[  T_{0},T_{1}\right]  $ & $\left[  0,3872\right]  $ & $1p_{\left[
T_{0},T_{1}\right]  }$ & $3456.1$\\
&  &  & $2p_{\left[  T_{0},T_{1}\right]  }$ & $3871.7$\\
&  &  &  & \\
&  &  &  & \\
& $\left[  T_{1},T_{2}\right]  $ & $\left[  3872,4080\right]  $ & $1p_{\left[
T_{1},T_{2}\right]  }$ & $4017.3$\\
&  &  &  &
\end{tabular}
\caption{Calculated $0^{+}(1^{++})$ charmonium masses from $V(r),$ generically
denoted by $M_{GSPM},$ in the first two energy regions specified by the
thresholds $\left[  T_{i-1},T_{i}\right]  $ and their masses $\left[
M_{T_{i-1}},M_{T_{i}}\right]  $. }%
\label{tabsp1++}%
\end{table}
\end{center}

By proceeding in the same way for higher energy regions and for different
quantum numbers we get the complete GSPM bound state spectrum. But before
listing the calculated spectral masses it may be illustrative to analyze the
effect produced by just one threshold. For this purpose we shall compare the
results obtained from the Cornell potential with the ones obtained from a GSPM
with only one threshold.

Let us consider again $0^{+}(1^{++})$ states and calculate the spectrum if
only the threshold $T_{1}$ (corresponding to $D^{0}\overline{D^{\ast}}%
^{0}(2007)$) is present. Then there will be two energy regions. In the first
one, $\left[  M_{T_{0}}=0,M_{T_{1}}=3872\text{ MeV}\right]  $, the potential
is given by (\ref{VT0T1}). Hence there are two bound states, $1p_{\left[
T_{0},T_{1}\right]  }$ and $2p_{\left[  T_{0},T_{1}\right]  },$ with the
masses previously calculated (see Table~\ref{tabsp1++}) which have been listed
again in Table~\ref{tabsp1++onethreshold}. In the second energy region,
$\left[  M_{T_{1}}=3872MeV,\infty\right[  $, the potential reads
\[
V_{\left[  3872,\infty\right[  }(r)=\left\{
\begin{array}
[c]{c}%
1174.8\text{ MeV\ \ \ \ \ \ \ \ }r\leq1.46\text{ fm}\\
\\
\sigma r-\frac{\chi}{r}\text{\ \ \ \ \ \ \ \ \ \ \ \ \ }r\geq1.46\text{ fm}%
\end{array}
\right.
\]
and has an infinite number of bound states. The masses of the two lowest
states in this energy region are listed in Table~\ref{tabsp1++onethreshold}.

\begin{center}
\begin{table}[ptb]%
\begin{tabular}
[c]{cc|cc}%
$%
\begin{array}
[c]{c}%
\text{GSPM}\\
\text{States}%
\end{array}
$ & $%
\begin{array}
[c]{c}%
M_{GSPM}\\
\text{MeV}%
\end{array}
$ & $%
\begin{array}
[c]{c}%
M_{Cor}\\
\text{MeV}%
\end{array}
$ & $%
\begin{array}
[c]{c}%
\text{Cornell}\\
\text{States}%
\end{array}
$\\\hline
&  &  & \\
$1p_{\left[  T_{0},T_{1}\right]  }$ & $3456.1$ & $3456.2$ & $1p$\\
$2p_{\left[  T_{0},T_{1}\right]  }$ & $3871.7$ & $3910.9$ & $2p$\\
&  &  & \\
$1p_{\left[  T_{1},\infty\right[  }$ & $4029.3$ &  & \\
&  &  & \\
$2p_{\left[  T_{1},\infty\right[  }$ & $4303.3$ & $4294.6$ & $3p$\\
&  &  &
\end{tabular}
\caption{Calculated $0^{+}(1^{++})$ charmonium masses up to $4400$ MeV when
only the first threshold is considered: $M_{GSPM}$. Masses from the Cornell
potential, $M_{Cor},$ are also shown for comparison. Conventional
spectroscopic notation has been used to denote the Cornell states. }%
\label{tabsp1++onethreshold}%
\end{table}
\end{center}

For the sake of comparison we calculate the Cornell spectrum in the same
energy interval (from $0$ to $4400$ MeV) from the same values of the
parameters $\sigma$, $\chi$ and $m_{c}$ given by (\ref{valpar}). The results
are also listed in Table~\ref{tabsp1++onethreshold}. We immediately realize
that in the spectral region considered there are four GSPM states $\left(
1p_{\left[  T_{0},T_{1}\right]  },2p_{\left[  T_{0},T_{1}\right]
},1p_{\left[  T_{1},\infty\right[  },2p_{\left[  T_{1},\infty\right[
}\right)  $ for only three Cornell states $\left(  1p,2p,3p\right)  $.
Moreover, the masses of the first and fourth GSPM states are quite the same as
the masses of the first and third Cornell states. Since the GSPM potential
differs from the Cornell one in the incorporation of the threshold
$T_{1}\equiv D^{0}\overline{D^{\ast}}^{0}(2007)$ we may interpret these
results by saying that the second $\left(  2p_{\left[  T_{0},T_{1}\right]
}\right)  $ and third $\left(  1p_{\left[  T_{1},\infty\right[  }\right)  $
GSPM states are effectively describing the mixing of the second Cornell
$\left(  2p\right)  $ state with the $D^{0}\overline{D^{\ast}}^{0}(2007)$
configuration. Therefore the effect of the threshold is the appearance of one
more spectral state (notice though that if the $2p$ Cornell state were farther
above the threshold the GSPM would not generate the $2p_{\left[  T_{0}%
,T_{1}\right]  }$ state).

\bigskip

\begin{figure}
[ptb]
\begin{center}
\includegraphics[
height=2.4396in,
width=3.659in
]%
{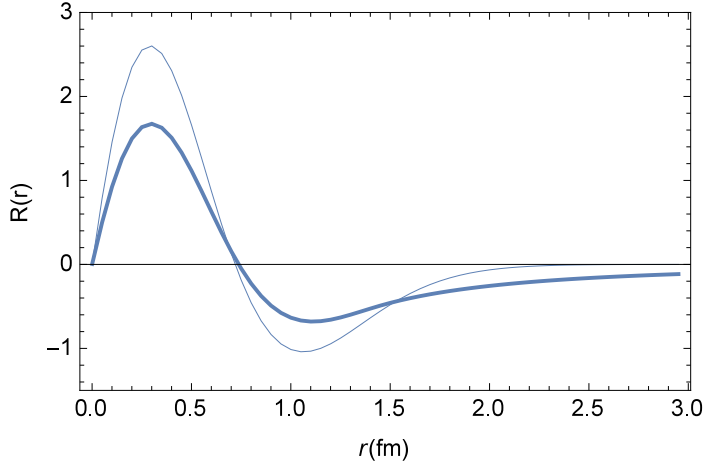}%
\caption{Radial wave functions R(r) (in units $fm^{-\frac{3}{2}}$) for the
$1^{++}\left(  2p_{\left[  T_{0},T_{1}\right]  }\right)  $ GSPM state (thick
line) and the $1^{++}\left(  2p\right)  $ Cornell state (thin line).}%
\label{Figx3872wf}%
\end{center}
\end{figure}

It may also be interesting to compare the resulting radial wave functions for
the $2p_{\left[  T_{0},T_{1}\right]  }$ GSPM state and the $2p$ Cornell state.
This comparison is drawn in Fig. \ref{Figx3872wf}.%

As can be checked the $2p_{\left[  T_{0},T_{1}\right]  }$ radial wave function
extends to much larger distances than the $2p$ one. If we consider the
$2p_{\left[  T_{0},T_{1}\right]  }$ state as an effective description of the
experimental $X(3872)$ and the $2p$ Cornell state as describing a (non
experimental) conventional $\chi_{c1}\left(  2p\right)  $ state then it is
clear the difference between them. The comparison of the respective root
square mean radii, $3.6$ fm for $X(3872)$ and $1.1$ fm for $\chi_{c1}\left(
2p\right)  $, indicates the screening of the heavy quark color charges in
$X(3872)$ due to the presence of the threshold. (Let us point out that in a
couple channel treatment involving quenched quark-antiquark and meson-meson
configurations this would correspond to the presence of a $D^{0}%
\overline{D^{\ast}}^{0}(2007)$ wave function component.)

\subsubsection{$0^{+}\left(  J^{++}\right)  $ GSPM States}

The spectrum for $0^{+}(J^{++})$ $c\overline{c}$ states from $V(r)$ is shown
in Table~\ref{tab4a}. The spectrum from the Cornell potential $V_{Cor}(r)$
given by (\ref{Cor}) with the same values of the parameters $\sigma$, $\chi$
and $m_{c}$ is also listed for comparison.

\begin{center}
\begin{table}[ptb]%
\begin{tabular}
[c]{cccccc}%
$J^{PC}$ & $%
\begin{array}
[c]{c}%
\text{GSPM}\\
\text{States}\\
k_{\left[  T_{i-1},T_{i}\right]  }%
\end{array}
$ & $%
\begin{array}
[c]{c}%
M_{GSPM}\\
\text{MeV}%
\end{array}
$ & $%
\begin{array}
[c]{c}%
M_{PDG}\\
\text{MeV}%
\end{array}
$ & $%
\begin{array}
[c]{c}%
M_{Cor}\\
\text{MeV}%
\end{array}
$ & $%
\begin{array}
[c]{c}%
\text{Cornell}\\
\text{States}\\
k
\end{array}
$\\\hline
&  &  &  &  & \\
$0^{++}$ & $1p_{\left[  T_{0},T_{1}\right]  }$ & $3456.1$ & $3414.75\pm0.31$ &
$3456.2$ & $1p$\\
$1^{++}$ & $1p_{\left[  T_{0},T_{1}\right]  }$ & $3456.1$ & $3510.66\pm0.07$ &
$3456.2$ & $1p$\\
$2^{++}$ & $1p_{\left[  T_{0},T_{1}\right]  }$ & $3456.1$ & $3556.20\pm0.09$ &
$3456.2$ & $1p$\\
&  &  &  &  & \\
$1^{++}$ & $2p_{\left[  T_{0},T_{1}\right]  }$ & $3871.7$ & $3871.69\pm0.17$ &
$3910.9$ & $2p$\\
&  &  &  &  & \\
$0^{++}$ & $1p_{\left[  T_{1},T_{2}\right]  }$ & $3897.9$ & $3918.4\pm1.9$ &
$3910.9$ & $2p$\\
&  &  &  &  & \\
$2^{++}$ & $2p_{\left[  T_{0},T_{1}\right]  }$ & $3903.0$ & $3927.2\pm2.6$ &
$3910.9$ & $2p$\\
&  &  &  &  & \\
$1^{++}$ & $1p_{\left[  T_{1},T_{2}\right]  }$ & $4017.3$ &  &  & \\
&  &  &  &  & \\
$0^{++}$ & $1p_{\left[  T_{3},T_{4}\right]  }$ & $4140.2$ &  &  & \\
&  &  &  &  & \\
&  &  & $X\left(  4140\right)  $ &  & \\
&  &  &  &  & \\
$2^{++}$ & $1p_{\left[  T_{1},T_{2}\right]  }$ & $4140.2$ &  &  & \\
&  &  &  &  & \\
&  &  &  &  & \\
&  &  &  &  & \\
$0^{++}$ & $1p_{\left[  T_{4},T_{5}\right]  }$ & $4325.1$ & $X\left(
4350\right)  $ & $4294.6$ & $3p$\\
&  &  &  &  &
\end{tabular}
\caption{Calculated $J^{++}$ charmonium masses from $V(r):M_{GSPM}$ up to
$4350$ MeV. The $0^{++}\left(  1p_{\left[  T_{2},T_{3}\right]  }\right)  $ row
has been omitted since there is no GSPM bound state in that energy region. For
$1^{++}$ we do not list any state above $4080$ MeV due to the current
incomplete knowledge about thresholds above this energy. The same for $2^{++}$
states above $4224$ MeV. Masses for experimental resonances, $M_{PDG},$ have
been taken from \cite{PDG14} (when a resonance appears in the Particle Listing
section of \cite{PDG14} but not in the Summary Table we write the name of the
resonance that contains the nominal mass between parenthesis). For $p$ waves
we quote separately the $np_{0}$, $np_{1}$ and $np_{2}$ states. Masses from
the Cornell potential, $M_{Cor},$ are also shown for comparison. }%
\label{tab4a}%
\end{table}
\end{center}

A glance at the table confirms the presence of a bigger number of GSPM states
than Cornell ones even ignoring possible additional $1^{++}$ states above
$4080$ MeV and $2^{++}$ states above $4224$ MeV. More precisely there are (at
least) four $J^{++}$ GSPM states in the energy interval $4000-4400$ MeV for
only one Cornell state. Since the calculated masses of three of these GSPM
states are in good correspondence with the masses of $X\left(  4140\right)  $
$0^{+}\left(  ?^{?+}\right)  $ and $X\left(  4350\right)  $ $0^{+}\left(
?^{?+}\right)  $, the currently existing experimental candidates to
$0^{+}(J^{++})$ states in that energy interval (see Particle Listing in
\cite{PDG14}; see also \cite{Ols14}), a tentative assignment of GSPM states to
these candidates has been done in Table~\ref{tab4a}. From it a guess for their
unknown quantum numbers comes out: $X\left(  4140\right)  $ $0^{+}\left(
0^{++}\right)  $ or $X\left(  4140\right)  $ $0^{+}\left(  2^{++}\right)  $
and $X\left(  4350\right)  $ $0^{+}\left(  0^{++}\right)  .$

\bigskip

Furthermore the model predicts the existence of at least two new $0^{+}%
(J^{++})$ resonances in the energy interval considered. One of them, that we
shall call $C(4140)$ ($C$ standing for theoretical candidate) would be
assigned to the $2^{++}\left(  1p_{\left[  T_{1},T_{2}\right]  }\right)  $ or
$0^{++}\left(  1p_{\left[  T_{3},T_{4}\right]  }\right)  $ GSPM state at
$4140.2$ MeV (see Table~\ref{tab4a}). Let us note that the existence of this
state in the GSPM is linked to the existence of the $0^{++}\left(  1p_{\left[
T_{3},T_{4}\right]  }\right)  $ or $2^{++}\left(  1p_{\left[  T_{1}%
,T_{2}\right]  }\right)  $ state that we have assigned to $X\left(
4140\right)  $; as both states are in between the same thresholds ($D^{\ast
}(2007)^{0}\overline{D^{\ast}}^{0}(2007)$ and $D_{s}^{\ast+}D_{s}^{\ast-}$)
the central potential used does not make any difference for $J=0$ and $J=2$.
The other new resonance that we shall call $C(4017)$ would be assigned to the
$1^{++}\left(  1p_{\left[  T_{1},T_{2}\right]  }\right)  $ GSPM state at
$4017.3$ MeV. As shown above this resonance is generated altogether with
$X(3872)$ as an effect of the introduction of the $D^{0}\overline{D^{\ast}%
}^{0}(2007)$ threshold. Hence the existence of $C(4017)$ seems to be
unavoidable if the mechanism proposed for the generation of $X(3872)$ is the
correct one.

\bigskip

For the sake of completeness let us mention that for energies quite below the
first thresholds the calculated spectrum is of Cornell type giving rise to
degenerate $J^{++}=\left(  0,1,2\right)  ^{++}$ states. This degeneracy is
broken for energies reaching the first thresholds (and beyond) due to the
different values of the threshold masses in each case.

\bigskip

We might then conclude that an assignment of GSPM states to the existing well
established or possible candidates to $0^{+}(J^{++})$ resonances is feasible.
With respect to the observed differences between the calculated GSPM masses
and data we shall assume that the experimental values can be reached from the
GSPM\ ones through perturbative corrections to the hamiltonian. The
experimental confirmation of the candidates and the discovery of the new
predicted resonances could give definite support to this conclusion.

\subsubsection{$0\left(  1^{--}\right)  $ GSPM States}

The spectrum for $0^{-}(1^{--})$ $c\overline{c}$ states from $V(r)$ up to
$4200$ MeV (quite below the first threshold located at $4287$ MeV, see
Table~\ref{tab1--}) is shown in Table~\ref{tab3}. The spectrum from the
Cornell potential $V_{Cor}(r)$ given by (\ref{Cor}) with the same values of
the parameters $\sigma$, $\chi$ and $m_{c}$ is also listed for comparison.
\begin{table}[ptb]%
\begin{tabular}
[c]{cccccc}%
$J^{PC}$ & $%
\begin{array}
[c]{c}%
\text{GSPM}\\
\text{States}\\
k_{\left[  T_{i-1},T_{i}\right]  }%
\end{array}
$ & $%
\begin{array}
[c]{c}%
M_{GSPM}\\
\text{MeV}%
\end{array}
$ & $%
\begin{array}
[c]{c}%
M_{PDG}\\
\text{MeV}%
\end{array}
$ & $%
\begin{array}
[c]{c}%
M_{Cor}\\
\text{MeV}%
\end{array}
$ & $%
\begin{array}
[c]{c}%
\text{Cornell}\\
\text{States}\\
k
\end{array}
$\\\hline
&  &  &  &  & \\
$1^{--}$ & $1s_{\left[  T_{0},T_{1}\right]  }$ & $3046.0$ & $3096.916\pm0.011$
& $3046.0$ & $1s$\\
& $2s_{\left[  T_{0},T_{1}\right]  }$ & $3632.2$ & $3686.09\pm0.04$ & $3632.2$
& $2s$\\
& $1d_{\left[  T_{0},T_{1}\right]  }$ & $3743.5$ & $3773.15\pm0.33$ & $3743.5$
& $1d$\\
& $3s_{\left[  T_{0},T_{1}\right]  }$ & $4063.2$ & $4039\pm1$ & $4065.8$ & $3s
$\\
& $2d_{\left[  T_{0},T_{1}\right]  }$ & $4139.3$ & $4191\pm5$ & $4142.8$ & $2d
$\\
&  &  &  &  &
\end{tabular}
\caption{Calculated $1^{--}$ charmonium masses from $V(r):M_{GSPM}.$ Masses
for experimental resonances, $M_{PDG},$ have been taken from \cite{PDG14}.
Masses from the Cornell potential, $M_{Cor},$ are also shown for comparison.}%
\label{tab3}%
\end{table}

An almost pure Cornell like spectrum (very little threshold effects) is
obtained in this energy region as can be checked by comparing the calculated
GSPM masses with the Cornell ones.

\bigskip

As explained before the GSPM\ can not be reliably applied to calculate the
masses of higher spectral states in this case. Nonetheless a qualitative
analysis of the possible mixing configuration content in some of the well
established higher spectral resonances can be carried out. Let us centre for
instance in $X(4260)$ lying close below the first (degenerate) threshold. Let
us examine whether this resonance could be obtained or not if only the
threshold $T_{11}\equiv D^{0}\overline{D_{1}}^{0}(2420)$ at $4287$ MeV were
present. Then the resulting GSPM spectrum from $4.0$ to $4.5$ GeV would be as
listed in Table~\ref{tab4260} where the Cornell spectrum is also given for comparison.

\bigskip

\begin{table}[ptb]%
\begin{tabular}
[c]{cc|cc}%
$%
\begin{array}
[c]{c}%
\text{GSPM}\\
\text{States}%
\end{array}
$ & $%
\begin{array}
[c]{c}%
M_{GSPM}\\
\text{MeV}%
\end{array}
$ & $%
\begin{array}
[c]{c}%
M_{Cor}\\
\text{MeV}%
\end{array}
$ & $%
\begin{array}
[c]{c}%
\text{Cornell}\\
\text{States}%
\end{array}
$\\\hline
&  &  & \\
$3s_{\left[  T_{0},\widetilde{T}_{1}\right]  }$ & $4063.2$ & $4065.8$ & $3s$\\
$2d_{\left[  T_{0},\widetilde{T}_{1}\right]  }$ & $4139.3$ & $4142.8$ & $2d$\\
&  &  & \\
$1s_{\left[  \widetilde{T}_{1},\infty\right[  }$ & $4337.3$ &  & \\
&  &  & \\
$1d_{\left[  \widetilde{T}_{1},\infty\right[  }$ & $4454.1$ & $4436.5$ &
$4s$\\
&  &  & \\
$2s_{\left[  \widetilde{T}_{1},\infty\right[  }$ & $4483.5$ & $4496.1$ & $3d$%
\end{tabular}
\caption{Calculated $1^{--}$ charmonium masses from $4.0$ to $5.0$ GeV when
only one non degenerate threshold $\widetilde{T}_{1}\equiv D^{0}%
\overline{D_{1}}^{0}(2420)$ is considered: $M_{GSPM}.$ Masses from the Cornell
potential, $M_{Cor},$ are also shown for comparison.}%
\label{tab4260}%
\end{table}A glance at the table shows that the presence of the threshold
would generate a new spectral state at $4337$ MeV as compared to the Cornell
case. But there would not be any chance to obtain a resonance close below
threshold, as the $X(4260)$. The explanation for this has to do with the fact
that there is not any Cornell state close below or above threshold from which
such resonance could be formed by the effect of the threshold.

The situation could change by considering the additional effect of the other
degenerate threshold $T_{12}\equiv D^{0}\overline{D_{1}}^{0}(2430)$ on the new
spectral state. As the mass of this state, $4337$ MeV, is close above $T_{12}%
$, it could be shifted down to a value below the threshold as experimentally
observed (notice that some additional attraction could also be provided by the
$D^{\ast}(2007)^{0}\overline{D_{0}^{\ast}}^{0}(2400)$ threshold due to its
large width). In this regard a refined version of the GSPM, incorporating a
lesser abrupt change in the potential when approaching the threshold, could
allow for a consistent treatment of the degenerate as well as the non
degenerate threshold effects through the different paths followed by the
potential to reach the threshold energy.

\bigskip

Therefore we might tentatively conclude (without any quantitative proof) that
the existence of $X(4260)$ could be related to the presence of degenerate
overlapping thresholds. Otherwise said $X(4260)$ could be the result of the
mixing of the quenched $c\overline{c}$ with $D^{0}\overline{D_{1}}^{0}(2420)$
and $D^{0}\overline{D_{1}}^{0}(2430)$ configurations.

\subsubsection{Electromagnetic Widths}

Electromagnetic decay rates of charmonium are sensitive to details of the wave
functions involved. Therefore their study might serve to test a quark model
and to discriminate it against others. One should realize though that when
ratios of decay rates are considered, similar results may be obtained from
different models. Indeed some of these ratios, involving transitions from
initial to final charmonium states, can be explained from heavy quark symmetry
considerations without reference to any particular dynamic model \cite{Cho95}.

Let us note that the GSPM assigns a differentiated state to each of the
existing non conventional experimental candidates to be a $0^{+}(J^{++})$
resonance. It also allows for an unambiguous assignment of states to
conventional $0^{+}(J^{++})$ and $0^{-}(1^{--})$ resonances below their first
thresholds. Therefore it can be consistently used for the analysis of
transitions involving these states.

\bigskip

We will focus on the calculation of electric dipole (E1) and two photon decay
widths for which a comparative analysis to data can be carried out. Thus, for
E1 decays we shall centre on transitions between spin triplet $P-$ wave and
$S-$ wave states for which the non relativistic E1 partial widths read
\cite{Eic80}
\begin{equation}
\Gamma_{E1}\left(  i\rightarrow f+\gamma\right)  =\frac{4\alpha e_{c}%
^{2}w_{if}^{3}\left(  2J_{f}+1\right)  }{27}\left\vert \left\langle
f\left\vert r\right\vert i\right\rangle \right\vert ^{2} \label{E1}%
\end{equation}
where $i$ and $f$ denote the initial (final) charmonium state, $\alpha$ stands
for the fine structure constant, $e_{c}=\frac{2}{3}$ is the charm quark
electric charge, $w_{if}$ is the photon energy
\begin{equation}
w_{if}=\frac{1}{2M_{i}}\left(  M_{i}^{2}-M_{f}^{2}\right)  \label{PhE}%
\end{equation}
and $\left\langle f\left\vert r\right\vert i\right\rangle $ is the dipole
matrix element
\begin{equation}
\left\langle f\left\vert r\right\vert i\right\rangle =%
{\displaystyle\int\limits_{0}^{\infty}}
R_{f}\left(  r\right)  r^{3}R_{i}\left(  r\right)  dr \label{dipme}%
\end{equation}
with $R_{f}\left(  r\right)  $ and $R_{i}\left(  r\right)  $ standing for the
radial wave functions of the final and initial state respectively.

From (\ref{E1}) we can easily establish the ratios:

$a)$
\begin{equation}
\frac{\Gamma_{E1}\left(  i\rightarrow f_{1}+\gamma\right)  }{\Gamma
_{E1}\left(  i\rightarrow f_{2}+\gamma\right)  }=\frac{w_{if_{1}}^{3}%
}{w_{if_{2}}^{3}}\frac{\left\vert \left\langle f_{1}\left\vert r\right\vert
i\right\rangle \right\vert ^{2}}{\left\vert \left\langle f_{2}\left\vert
r\right\vert i\right\rangle \right\vert ^{2}} \label{iratio}%
\end{equation}
for the case in which the same initial state decays into two final $\left(
f_{1}\text{ and }f_{2}\right)  $ states with the same value of $J_{f}$ and

$b)$
\begin{equation}
\frac{\Gamma_{E1}\left(  i_{1}\rightarrow f+\gamma\right)  }{\Gamma
_{E1}\left(  i_{2}\rightarrow f+\gamma\right)  }=\frac{w_{i_{1}f}^{3}%
}{w_{i_{2}f}^{3}}\frac{\left\vert \left\langle f\left\vert r\right\vert
i_{1}\right\rangle \right\vert ^{2}}{\left\vert \left\langle f\left\vert
r\right\vert i_{2}\right\rangle \right\vert ^{2}} \label{fratio}%
\end{equation}
for the case in which two initial states $\left(  i_{1}\text{ and }%
i_{2}\right)  $ decay into the same final state.

\bigskip

As for two photon transitions we shall consider the decays from $^{3}P_{0}$
and $^{3}P_{2}$ states. In the nonrelativistic limit the decay widths can be
expressed as \cite{Kwo88}
\begin{equation}
\Gamma\left(  i\left(  ^{3}P_{0}\right)  \rightarrow\gamma\gamma\right)
=\frac{27\alpha^{2}e_{c}^{4}}{m_{c}^{4}}\left\vert R_{i\left(  ^{3}%
P_{0}\right)  }^{^{\prime}}\left(  0\right)  \right\vert ^{2} \label{TPD3P0}%
\end{equation}%
\begin{equation}
\Gamma\left(  i\left(  ^{3}P_{2}\right)  \rightarrow\gamma\gamma\right)
=\frac{36\alpha^{2}e_{c}^{4}}{5m_{c}^{4}}\left\vert R_{i\left(  ^{3}%
P_{2}\right)  }^{^{\prime}}\left(  0\right)  \right\vert ^{2} \label{TPD3P2}%
\end{equation}
where $R_{i}^{^{\prime}}\left(  0\right)  $ stands for the derivative of the
radial wave function at the origin.

First order QCD radiative corrections to (\ref{TPD3P0}) and (\ref{TPD3P2}), in
the form of multiplying factors, have been calculated. For the effective value
of $\alpha_{s}=0.38$ we are using they are significant. This poses the need to
calculate them to higher order. Instead we shall keep the zeroth order
expressions to get a first approach to data and we shall use for practical
purposes the ratios%
\begin{equation}
\frac{\Gamma\left(  i_{2}\left(  ^{3}P_{0}\right)  \rightarrow\gamma
\gamma\right)  }{\Gamma\left(  i_{1}\left(  ^{3}P_{0}\right)  \rightarrow
\gamma\gamma\right)  }=\frac{\left\vert R_{i_{2}\left(  ^{3}P_{0}\right)
}^{^{\prime}}\left(  0\right)  \right\vert ^{2}}{\left\vert R_{i_{1}\left(
^{3}P_{0}\right)  }^{^{\prime}}\left(  0\right)  \right\vert ^{2}}
\label{3P02F}%
\end{equation}%
\begin{equation}
\frac{\Gamma\left(  i_{2}\left(  ^{3}P_{2}\right)  \rightarrow\gamma
\gamma\right)  }{\Gamma\left(  i_{1}\left(  ^{3}P_{2}\right)  \rightarrow
\gamma\gamma\right)  }=\frac{\left\vert R_{i_{2}\left(  ^{3}P_{2}\right)
}^{^{\prime}}\left(  0\right)  \right\vert ^{2}}{\left\vert R_{i_{1}\left(
^{3}P_{2}\right)  }^{^{\prime}}\left(  0\right)  \right\vert ^{2}}
\label{3P22F}%
\end{equation}
where the multiplying factors cancel out.

\bigskip

\bigskip

\begin{center}
$\mathbf{\chi}_{cJ}\mathbf{(1P)}$
\end{center}

\bigskip

The $\chi_{cJ}\left(  1P\right)  $ resonances are identified with the
$J^{++}\left(  1p_{\left[  T_{0},T_{1}\right]  }\right)  $ GSPM states which
are practically identical to the $J^{++}\left(  1p\right)  $ Cornell states.
As the $J/\psi$ description is also the same with both models they give the
same results for the $\chi_{cJ}\left(  1P\right)  \rightarrow\gamma J/\psi$
decay widths if the same values for the photon energies are chosen.

The calculated GSPM widths are shown in Table~\ref{E1cJ(1P)} where the dipole
matrix elements are also tabulated. For the photon energies the experimental
values have been used. This can be justified under our former assumption that
the experimental masses can be reached from the GSPM\ ones through first order
perturbative corrections to the hamiltonian (let us remind that no
modification of the wave functions is then generated).

\begin{center}
\begin{table}[ptb]%
\begin{tabular}
[c]{cccccc}%
$f$ & $i$ & $%
\begin{array}
[c]{c}%
\left(  w_{if}\right)  _{Exp}\\
\text{MeV}%
\end{array}
$ & $%
\begin{array}
[c]{c}%
\left\vert \left\langle f\left\vert r\right\vert i\right\rangle \right\vert
^{2}\\
\text{fm}^{2}%
\end{array}
$ & $%
\begin{array}
[c]{c}%
\left(  \Gamma_{E1}\left(  i\rightarrow f+\gamma\right)  \right)  _{GSPM}\\
\text{MeV}%
\end{array}
$ & $%
\begin{array}
[c]{c}%
\left(  \Gamma_{E1}\left(  i\rightarrow f+\gamma\right)  \right)  _{Exp}\\
\text{MeV}%
\end{array}
$\\\hline
&  &  &  &  & \\
$J/\psi$ & $\chi_{c0}\left(  1P\right)  $ & $303.04$ & $0.198$ & $0.20$ &
$0.13\pm0.02$\\
&  &  &  &  & \\
& $\chi_{c1}\left(  1P\right)  $ & $389.36$ & $0.198$ & $0.43$ & $0.29\pm
0.02$\\
&  &  &  &  & \\
& $\chi_{c2}\left(  1P\right)  $ & $429.63$ & $0.198$ & $0.58$ & $0.37\pm
0.04$\\
&  &  &  &  &
\end{tabular}
\caption{Calculated E1 dipole matrix elements (fourth column) and decay widths
(fifth column) for $\chi_{cJ}\left(  1P\right)  \rightarrow\gamma J/\psi$.
Photon energies (third column) from (\ref{PhE}) with experimental masses
\cite{PDG14}. Widths data (sixth column) from \cite{PDG14}. }%
\label{E1cJ(1P)}%
\end{table}
\end{center}

As can be checked the values obtained are 30\% off the experimental intervals.
This can be considered a reasonable first approach to data and a starting
point to include additional corrections (see \cite{Eic08} and references
therein). Although we do not proceed here along this line it is worth to point
out that the central values of the experimental ratios
\begin{equation}
\frac{\left(  \Gamma_{E1}\left(  \chi_{c2}\left(  1P\right)  \rightarrow\gamma
J/\psi\right)  \right)  _{Exp}}{\left(  \Gamma_{E1}\left(  \chi_{c0}\left(
1P\right)  \rightarrow\gamma J/\psi\right)  \right)  _{Exp}}=2.9\pm0.8
\label{ExpE120}%
\end{equation}
and
\begin{equation}
\frac{\left(  \Gamma_{E1}\left(  \chi_{c1}\left(  1P\right)  \rightarrow\gamma
J/\psi\right)  \right)  _{Exp}}{\left(  \Gamma_{E1}\left(  \chi_{c0}\left(
1P\right)  \rightarrow\gamma J/\psi\right)  \right)  _{Exp}}=2.2\pm0.6
\label{EXPE121}%
\end{equation}
are in good agreement with the experimental photon energy ratios
\[
\frac{\left(  w_{\chi_{c2\left(  1P\right)  J/\psi}}^{3}\right)  _{Exp}%
}{\left(  w_{\chi_{c0\left(  1P\right)  J/\psi}}^{3}\right)  _{Exp}}=2.9
\]%
\[
\frac{\left(  w_{\chi_{c2\left(  1P\right)  J/\psi}}^{3}\right)  _{Exp}%
}{\left(  w_{\chi_{c1\left(  1P\right)  J/\psi}}^{3}\right)  _{Exp}}=2.1
\]
Taking into account (\ref{fratio}) this suggests the additional corrections
should not introduce any significant difference among the $\chi_{c0}\left(
1P\right)  ,$ $\chi_{c1}\left(  1P\right)  $ and $\chi_{c2}\left(  1P\right)
$ wave functions.

\bigskip

\bigskip

Regarding two photon decay widths, the degeneracy of the $J^{++}\left(
1p_{\left[  T_{0},T_{1}\right]  }\right)  $ GSPM states gives rise from
(\ref{TPD3P0}) and (\ref{TPD3P2}) to the ratio%
\[
\frac{\left(  \Gamma\left(  \chi_{c0}\left(  1P\right)  \rightarrow
\gamma\gamma\right)  \right)  _{GSPM}}{\left(  \Gamma\left(  \chi_{c2}\left(
1P\right)  \rightarrow\gamma\gamma\right)  \right)  _{GSPM}}=\frac{27}{\left(
\frac{36}{5}\right)  }=3.75
\]
within the experimental interval \cite{PDG14}
\[
\frac{\left(  \Gamma\left(  \chi_{c0}\left(  1P\right)  \rightarrow
\gamma\gamma\right)  \right)  _{Exp}}{\left(  \Gamma\left(  \chi_{c2}\left(
1P\right)  \rightarrow\gamma\gamma\right)  \right)  _{Exp}}=\frac
{2.3\pm0.4\text{ KeV}}{0.53\pm0.07\text{ KeV}}=4.3\pm1.6
\]

\bigskip

\bigskip

\bigskip

\begin{center}
$\mathbf{\chi}_{c2}\left(  \mathbf{2P}\right)  $

\bigskip
\end{center}

\begin{figure}
[ptb]
\begin{center}
\includegraphics[
height=2.4396in,
width=3.659in
]%
{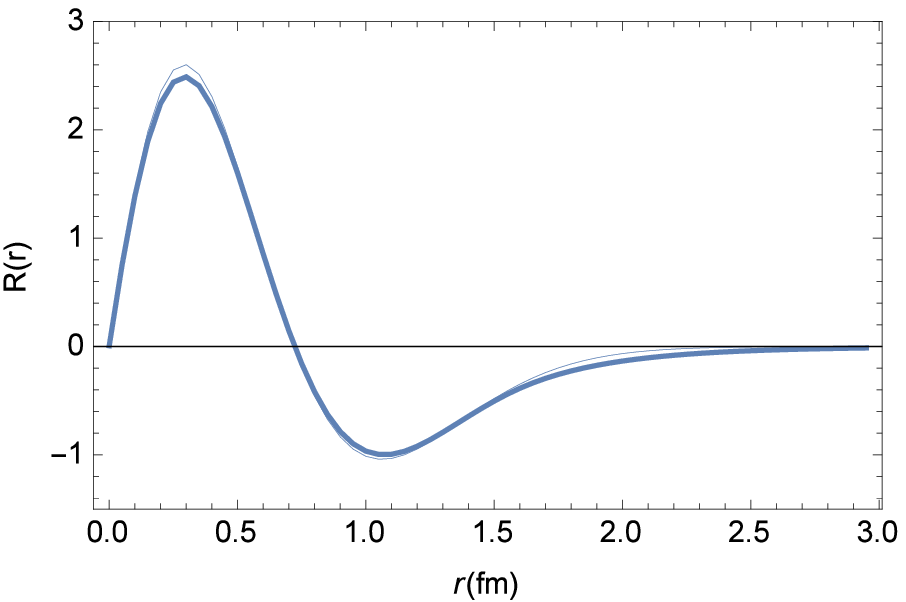}%
\caption{Radial wave functions R(r) (in units $fm^{-\frac{3}{2}}$) for the
$2^{++}\left(  2p_{\left[  T_{0},T_{1}\right]  }\right)  $ GSPM state (thick
line) and the $2^{++}\left(  2p\right)  $ Cornell state (thin line).}%
\label{Figchic2_2pwf}%
\end{center}
\end{figure}

The $\chi_{c2}\left(  2P\right)  $ resonance at $3927$ MeV is assigned to the
$2^{++}\left(  2p_{\left[  T_{0},T_{1}\right]  }\right)  $ GSPM state that
differs little from the $2^{++}\left(  2p\right)  $ Cornell one, as shown in
Fig. \ref{Figchic2_2pwf}.%

Only an experimental lower bound for the two photon decay width is known
\cite{PDG14}
\[
\left(  \Gamma\left(  \chi_{c2}\left(  2p\right)  \rightarrow\gamma
\gamma\right)  \right)  _{Exp}>0.21\pm0.04\text{ eV}%
\]
from%
\[
\left(  \Gamma\left(  \chi_{c2}\left(  2p\right)  \rightarrow\gamma
\gamma\right)  \mathcal{B}\left(  \chi_{c2}\left(  2p\right)  \rightarrow
D\overline{D}\right)  \right)  _{Exp}=0.21\pm0.04\text{ eV}%
\]
where $\mathcal{B}\ $stands for branching fraction.

\bigskip

From the calculated GSPM\ wave functions we get from (\ref{3P22F}) the ratio%
\[
\frac{\left(  \Gamma\left(  \chi_{c2}\left(  2p\right)  \rightarrow
\gamma\gamma\right)  \right)  _{GSPM}}{\left(  \Gamma\left(  \chi_{c2}\left(
1p\right)  \rightarrow\gamma\gamma\right)  \right)  _{GSPM}}=\frac{\left(
\left\vert R_{2^{++}\left(  2p_{\left[  T_{0},T_{1}\right]  }\right)
}^{^{\prime}}\left(  0\right)  \right\vert ^{2}\right)  _{GSPM}}{\left(
\left\vert R_{2^{++}\left(  1p_{\left[  T_{0},T_{1}\right]  }\right)
}^{^{\prime}}\left(  0\right)  \right\vert ^{2}\right)  _{GSPM}}=1.34
\]
By assuming that this value is a reasonable approach to the experimental ratio
we might expect the approximated values
\[
\Gamma\left(  \chi_{c2}\left(  2p\right)  \rightarrow\gamma\gamma\right)
\simeq1.34\left(  \Gamma\left(  \chi_{c2}\left(  1p\right)  \rightarrow
\gamma\gamma\right)  \right)  _{Exp}=0.71\pm0.09\text{ keV}%
\]%
\[
\mathcal{B}\left(  \chi_{c2}\left(  2p\right)  \rightarrow D\overline
{D}\right)  \simeq0.30\pm0.10
\]

\bigskip

\bigskip

\begin{center}
$\mathbf{X(3872)}$

\bigskip
\end{center}

As shown before the $1^{++}\left(  2p_{\left[  T_{0},T_{1}\right]  }\right)  $
GSPM state is identified with the $X(3872)$ whose mass has been used to fine
tune the charm quark mass.

Concerning electromagnetic decays the ratio
\[
A\equiv\frac{\Gamma\left(  X(3872)\rightarrow\gamma\psi\left(  2s\right)
\right)  }{\Gamma\left(  X(3872)\rightarrow\gamma J/\psi\right)  }%
\]
has been recently measured \cite{LHCb} to be%
\[
A_{Exp}=2.46\pm0.64\pm0.29
\]
compatible with the previous value $3.4\pm1.4$ \cite{Bab09} and the upper
bound $<2.1$ \cite{Bel11}.

From (\ref{iratio}) the GSPM\ gives for this ratio the value
\[
A_{GSPM}=2.01
\]
calculated from the dipole matrix elements%
\[
\left\vert \left\langle \psi\left(  2s\right)  \left\vert r\right\vert
X(3872)\right\rangle _{GSPM}\right\vert ^{2}=0.2856\text{ fm}^{2}%
\]%
\[
\left\vert \left\langle J/\psi\left\vert r\right\vert X(3872)\right\rangle
_{GSPM}\right\vert ^{2}=0.0025\text{ fm}^{2}%
\]
and the experimental values of the photon energies%
\[
\left(  w_{X(3872)\psi\left(  2s\right)  }\right)  _{Exp}=181.25\text{ MeV}%
\]%
\[
\left(  w_{X(3872)J/\psi}\right)  _{Exp}=697.19\text{ MeV}%
\]
Therefore a full compatibility with existing data comes out. We should point
out though that quite the same result would be obtained for the dipole matrix
elements by using the $1^{++}\left(  2p\right)  $ Cornell state wave function
instead of the $1^{++}\left(  2p_{\left[  T_{0},T_{1}\right]  }\right)  $ GSPM
one. As the main difference between these two wave functions is the long tail
of $2p_{\left[  T_{0},T_{1}\right]  }$ state as compared to that of $2p$ (see
Fig. \ref{Figx3872wf}) we may conclude that these radiative decays are not
sensitive to the long distance nature of $X(3872)$. This can be understood by
the negligible long distance overlap of the $1^{++}\left(  2p_{\left[
T_{0},T_{1}\right]  }\right)  $ state with $J/\psi$ and $\psi\left(
2s\right)  .$ (The calculated root mean square radii for $J/\psi$ and
$\psi\left(  2s\right)  $ are respectively $0.5$ fm and $0.9$ fm.) It should
be mentioned that the same conclusion has been also inferred by other authors
using molecular descriptions for $X(3872)$ \cite{Meh11,Guo15}.

\bigskip

\bigskip

\bigskip

\begin{center}
$\mathbf{X(3915)}$

\bigskip
\end{center}

The Review of Particle Properties \cite{PDG14} has identified the $X(3915)$
with a conventional $\chi_{c0}\left(  2p\right)  $, this is with a
$0^{++}\left(  2p\right)  $ Cornell like state. This identification has been
criticized by some authors \cite{Guo12,ols14}. A major criticism is the lack
of evidence of $X(3915)\rightarrow D\overline{D}$ decays. From our estimation
above for the branching fraction $\mathcal{B}\left(  \chi_{c2}\left(
2p\right)  \rightarrow D\overline{D}\right)  \simeq0.30$ a similar result
could be expected for $\mathcal{B}\left(  \chi_{c0}\left(  2p\right)
\rightarrow D\overline{D}\right)  $ since the $0^{++}\left(  2p\right)  $ and
$2^{++}\left(  2p\right)  $ Cornell states are degenerate and the measured
values of the masses and total widths of $\chi_{c2}\left(  2p\right)  $ and
$X(3915)$ are quite similar.

\bigskip

Regarding electromagnetic processes a lower bound for the two photon decay
width \cite{PDG14}
\[
\left(  \Gamma\left(  X(3915)\rightarrow\gamma\gamma\right)  \right)
_{Exp}>54\pm9\text{ eV}%
\]
is known from
\[
\left(  \Gamma\left(  X(3915)\rightarrow\gamma\gamma\right)  \mathcal{B}%
\left(  X(3915)\rightarrow J/\psi\omega\right)  \right)  _{Exp}>54\pm9\text{
eV}%
\]

\bigskip

\begin{figure}
[ptb]
\begin{center}
\includegraphics[
height=2.4396in,
width=3.659in
]%
{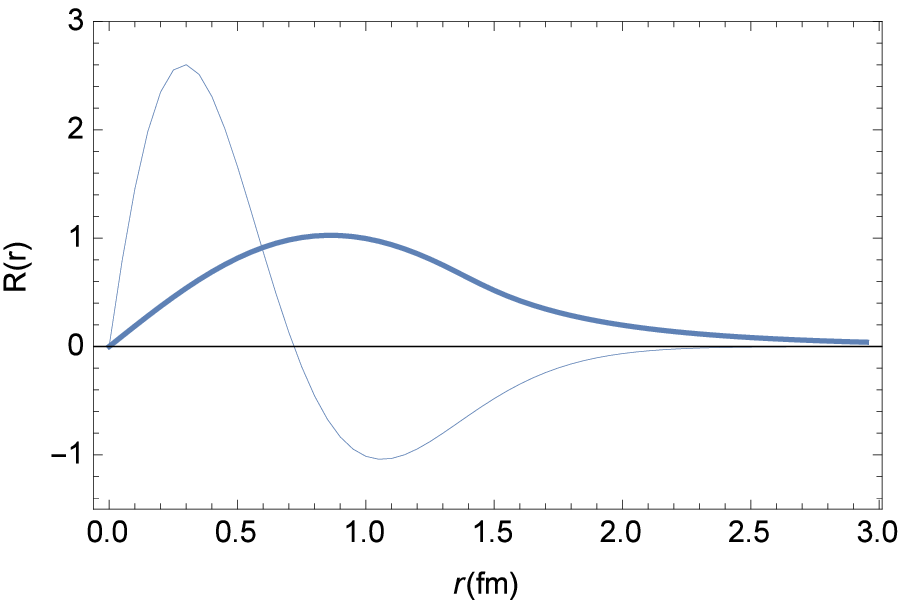}%
\caption{Radial wave functions R(r) (in units $fm^{-\frac{3}{2}}$) for the
$0^{++}\left(  1p_{\left[  T_{0},T_{1}\right]  }\right)  $ GSPM state (thick
line) and the $0^{++}\left(  2p\right)  $ Cornell state (thin line).}%
\label{Figx3915wf}%
\end{center}
\end{figure}

The GSPM assigns the $X(3915)$ to the $0^{++}\left(  1p_{\left[  T_{1}%
,T_{2}\right]  }\right)  $ state that differs greatly from the $0^{++}\left(
2p\right)  $ Cornell one as shown in Fig. $\ref{Figx3915wf}$.%

From the calculated GSPM\ wave functions we get from (\ref{3P02F}) the ratio%
\[
\frac{\left(  \Gamma\left(  X(3915)\rightarrow\gamma\gamma\right)  \right)
_{GSPM}}{\left(  \Gamma\left(  \chi_{c0}\left(  1P\right)  \rightarrow
\gamma\gamma\right)  \right)  _{GSPM}}=\frac{\left\vert R_{0^{++}\left(
1p_{\left[  T_{1},T_{2}\right]  }\right)  }^{^{\prime}}\left(  0\right)
\right\vert ^{2}}{\left\vert R_{0^{++}\left(  1p_{\left[  T_{0},T_{1}\right]
}\right)  }^{^{\prime}}\left(  0\right)  \right\vert ^{2}}=0.02
\]
Assuming again that this value is a reasonable approach to the experimental
ratio we might expect
\[
\Gamma\left(  X(3915)\rightarrow\gamma\gamma\right)  \simeq0.02\left(
\Gamma\left(  \chi_{c0}\left(  1p\right)  \rightarrow\gamma\gamma\right)
\right)  _{Exp}=44\pm7\text{ eV}%
\]
By combining this result with the experimental lower bound given above we
would get
\[
\Gamma\left(  X(3915)\rightarrow\gamma\gamma\right)  \simeq48\pm3\text{ eV}%
\]%
\[
\mathcal{B}\left(  X(3915)\rightarrow J/\psi\omega\right)  >0.88
\]

\bigskip

It should be emphasized that the identification of $X(3915)$ with
$0^{++}\left(  2p\right)  $ would give a completely different ratio%
\[
\frac{\left(  \Gamma\left(  X(3915)\rightarrow\gamma\gamma\right)  \right)
_{Cornell}}{\left(  \Gamma\left(  \chi_{c0}\left(  1P\right)  \rightarrow
\gamma\gamma\right)  \right)  _{Cornell}}=\frac{\left\vert R_{0^{++}\left(
2p\right)  }^{^{\prime}}\left(  0\right)  \right\vert ^{2}}{\left\vert
R_{0^{++}\left(  1p\right)  }^{^{\prime}}\left(  0\right)  \right\vert ^{2}%
}=1.2
\]
and consequently completely different values for the two photon decay width
$\left(  2640\text{ eV}\right)  $ and the branching fraction to $J/\psi\omega$
$\left(  >0.02\right)  .$

\bigskip

Certainly the big GSPM branching fraction for $X(3915)\rightarrow J/\psi
\omega$, OZI suppressed in the Cornell model, should be somehow justified. A
quantitative justification, if possible, would imply the development of a
strong decay theory above threshold within the GSPM framework which is outside
the scope of this work. Hence we shall limit here to a merely speculative
qualitative comment. Let us imagine for instance that due to the threshold
modulation the GSPM interaction favored, at the energy of $X(3915)$ and
through light quark pair creation out of the vacuum, the formation of color
octets made of heavy quark-light antiquark and viceversa. Then it would be
possible to have a dominant decay through reordering of the quarks in such a
state like $J/\psi\omega.$ On the contrary the Cornell states are known to
favor the formation of color singlets giving rise to the dominant fall apart
decay mode $D\overline{D}$.

\bigskip

Therefore the GSPM and Cornell descriptions represent incompatible scenarios
for the understanding of $X(3915).$ According to our analysis more detailed
data could definitely clarify the situation about the true nature (non
conventional or conventional) of this resonance.

\bigskip

\section{Summary\label{SIV}}

A nonrelativistic quark model called Generalized Screened Potential Model, or
abbreviate GSPM, previously used to calculate the bottomonium spectrum has
been applied to charmonium.

The model, whose interaction potential has a Cornell form but modulated by
meson-meson thresholds, has been used to calculate $0^{+}(J^{++})$ charmonium
masses up to $4.4$ GeV, a limit imposed to the application of the model by the
incomplete current knowledge of open charm meson-meson thresholds. As it
turned out to be the case in bottomonium a richer spectrum (bigger number of
bound states) than the one resulting from the non-screened Cornell potential
is predicted. However, differing from bottomonium where the lack of
data\ prevented the verification or refutationof such a prediction, there
exist in charmonium well established as well as candidates to non conventional
resonances in the energy interval analyzed. As a matter of fact the well
established $X(3872)$ is nicely described as a GSPM state that can be
interpreted as being generated from the $D^{0}\overline{D^{\ast}}^{0}(2007)$
threshold and the $1^{++}\left(  2p\right)  $ Cornell state. Regarding the
experimental candidates $X\left(  4140\right)  $ $0^{+}\left(  ?^{?+}\right)
$ and $X\left(  4350\right)  $ $0^{+}\left(  ?^{?+}\right)  $ a good spectral
correspondence with GSPM states is observed. Furthermore two new $0^{+}%
(J^{++})$ resonances are predicted, a $0^{+}(2^{++})$ or $0^{+}(0^{++})$ one
with mass around $4140$ MeV and a $0^{+}(1^{++})$ one with mass around $4017$
MeV. (Notice though that all the calculated GSPM masses, except for $X(3872)$
which is used to fine tune the quark mass, are below the experimental ones
what suggests that the masses of these new resonances could also be
underestimated.) The generation of these new resonances in the GSPM is related
to the presence of the states assigned to $X(4140)$ and $X(3872)$. Therefore
their discovery would constitute a definite check of the GSPM as a model for
the spectral description.

\bigskip

The GSPM has also been employed to evaluate the $0^{-}(1^{--})$ spectrum up to
$4.2$ GeV, a limit imposed to the applicability of the model by the presence
of degenerate and overlapping thresholds. The resulting spectrum in this
energy region is very much of Cornell type. In order to go further in energy
the model should be refined. With respect to this a simplified qualitative
analysis of the possible generation of $X\left(  4260\right)  $ seems to point
out that the joint effect from overlapping meson-meson configurations should
be an essential ingredient to be incorporated.

\bigskip

A study of electromagnetic decays of $0^{+}(J^{++})$ resonances, specifically
E1 and two photon decays for which there are data available, has also been
carried out. The calculated GSPM widths are fully compatible with existing
data. However, more detailed data are needed to perform a stringent check of
the GSPM. In this regard a thorough experimental analysis of $X(3915)$ is of
particular interest given the very different description coming out from the
GSPM\ and the Cornell models.

\bigskip

This work has been supported by Ministerio de Econom\'{\i}a y Competitividad
of Spain (MINECO) grant FPA2013-47443-C2-1-P, and by PrometeoII/2014/066 from
Generalitat Valenciana.

\end{document}